\documentclass[journal=jacsat,manuscript=article]{achemso}

\usepackage[version=3]{mhchem} 
\usepackage{amsmath,amssymb,amsfonts} 
\usepackage{multirow} 
\usepackage{color} 
\usepackage{soul}

\author{Asuka Hosokawa}
\affiliation[Tokyo University of Agriculture and Technology]
{Department of Mechanical Systems Engineering, Tokyo University of Agriculture and Technology, Naka-cho 2-24-16, Koganei, Tokyo, 184-8588, Japan}
\author{Kyota Kamamoto}
\affiliation[Tokyo University of Agriculture and Technology]
{Department of Mechanical Systems Engineering, Tokyo University of Agriculture and Technology, Naka-cho 2-24-16, Koganei, Tokyo, 184-8588, Japan}
\author{Hiroya Watanabe}
\affiliation[Tokyo University of Agriculture and Technology]
{Department of Mechanical Systems Engineering, Tokyo University of Agriculture and Technology, Naka-cho 2-24-16, Koganei, Tokyo, 184-8588, Japan}
\author{Hiroaki Kusuno}
\affiliation[Tokyo University of Agriculture and Technology]
{Department of Mechanical Systems Engineering, Tokyo University of Agriculture and Technology, Naka-cho 2-24-16, Koganei, Tokyo, 184-8588, Japan}
\author{Kazuya U. Kobayashi}
\affiliation[Nippon Institute of Technology]
{Department of Mechanical Engineering, Nippon Institute of Technology, Gakuendai 4-1, Miyashiro-machi, Minamisaitama-gun, Saitama, 345-8501, Japan}
\author{Yoshiyuki Tagawa}
\affiliation[Tokyo University of Agriculture and Technology]
{Department of Mechanical Systems Engineering, Tokyo University of Agriculture and Technology, Naka-cho 2-24-16, Koganei, Tokyo, 184-8588, Japan}
\email{tagawayo@cc.tuat.ac.jp}

\title[An \textsf{achemso} demo]
  {A phase diagram of the pinch-off behavior of impulsively-induced viscoelastic liquid jets}

\abbreviations{IR,NMR,UV}

\begin{document}

\begin{tocentry}
\begin{center}
	\includegraphics[width=0.76\textwidth]{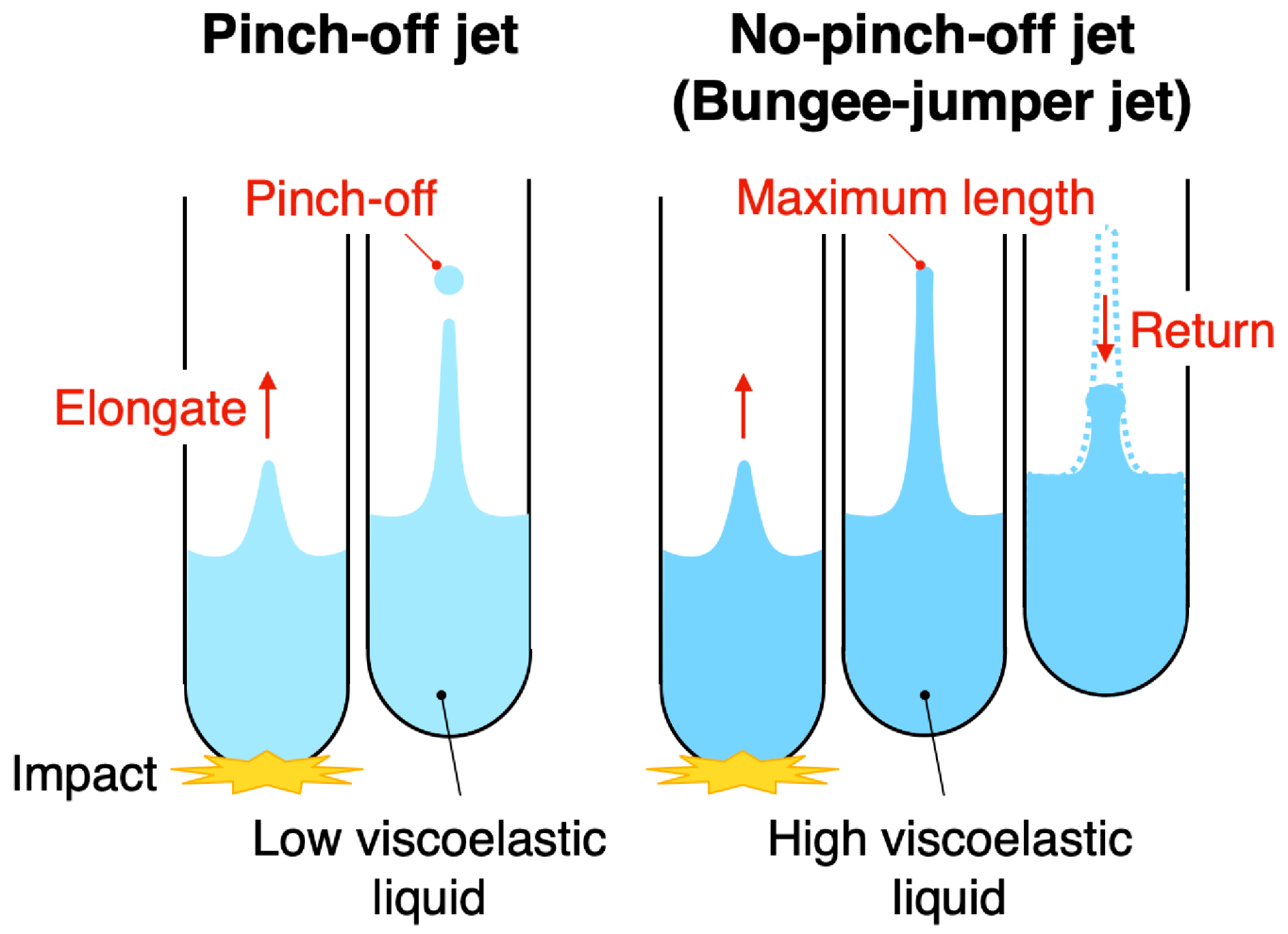}
\end{center}
\end{tocentry}

\begin{abstract}
 In this study, we systematically investigate the behaviors of viscoelastic liquid jets using an impulsive force, particularly in the high velocity and high elasticity regimes. The resulting jets are categorized into two types: (i) pinch-off jets, which break up during elongation after ejection, and (ii) no-pinch-off jets, which either retract to the nozzle after maximum elongation, known as `bungee-jumper jets' or return without elongation after ejection. We then propose criteria to delineate these regions using Reynolds number $Re$ and Weissenberg number $Wi$,  reflecting the initial conditions at the jet ejection and the solution's rheological properties, respectively. We find that pinch-off jets occur at $Re \gtrsim 23.4Wi$ in high elasticity regimes ($Wi \gtrsim 10$), and at $Re \gtrsim 250$ in low elasticity regimes ($Wi \lesssim 10$). In addition, we demonstrate that the phase diagram of these behaviors can be rationalized through the focused jet modeling using the finitely extensible non-linear elastic dumbbell model with the Chilcott–Rallison closure approximation (FENE-CR).

\end{abstract}

\section{Introduction}
Liquid jets have been studied in the past from a variety of perspectives due to their significance in fundamental research and their widespread occurrence\citep{eggers2008}. A typical liquid jet is a focused jet in which the diameter of the jet tip is smaller than the diameter of the jet bottom\cite{tagawa2012,peters2013,kiyama2016}. Focused jets are caused by solid spheres or droplets impinging on the surface of a liquid bath \cite{worthington1908} or by the formation of standing waves (Faraday waves) on the surface of a liquid bath\citep{zeff2000}. It can also be caused by rapid acceleration of a concave gas-liquid interface by an impulsive force\citep{Antkowiak2007}. Such impulsively-induced focused jets are expected to be applied to needle-free injection of drugs \cite{Joy2006} and inkjet printing technology \cite{onuki2018} because of their thin tips \citep{gordillo2020} and high speed.\citep{tagawa2012}

For inkjet printing applications, it is very important to understand the mechanisms of jet formation and breakup. The formation and breakup of Newtonian fluid jets is well known and predictable by considering Rayleigh-Plateau instability.\citep{Zhang2022} After pinch-off from the nozzle, the jet forms a single droplet or multiple droplets (satellite).\citep{Detlef2022} Since satellite generates unwanted droplets on the target substrate and degrade print quality, it is desirable for a single droplet to be formed during actual printing. Therefore, methods to generate a single droplet have been investigated. 

One method is to increase the viscosity of the liquid. Delrot et al. \citep{Delrot2016} showed that the condition for single droplet formation is expressed as $0.1<Oh<1$ using the Ohnesorge number, $Oh = \mu / \sqrt{\rho r \sigma}$, where $\mu$ is the liquid viscosity, $\rho$ is the liquid density, $r$ is the characteristic length scale for the flow of interest, and $\sigma$ is the liquid–air surface tension. Another way is to add polymers to the liquid to give it viscoelastic properties. Viscoelastic liquid jets exhibit a characteristic phenomenon called `beads-on-a-string', in which droplets and strings of liquid form a series of beads, suppressing breakup.\citep{clasen2006} Previous studies have shown that the addition of small amounts of high molecular weight polymers to liquids suppresses satellite formation.\citep{mcilroy2013} Conversely, when large amounts of high molecular weight polymers are added, elasticity can delay or prevent jet formation from the nozzle. The characteristic jet produced in this case is called a `bungee-jumper jet', which extends to its maximum length after formation and then gets pulled back to the gas-liquid interface without pinch-off.\citep{morrison2010}
In order to determine the ideal amount of polymer to add, jet behavior has been investigated by varying the molecular weight and concentration of polymers in drop-on-demand (DoD) printing, a common inkjet technology. Jet behavior has been classified by experiments\citep{bazilevskii2005,hoath2012,mcilroy2013,sen2021} and simulations,\citep{morrison2010,turkoz2021} mainly using polymer concentration and molecular weight. Physical considerations have also been made on the influence of viscoelasticity on jet behavior by modeling the phenomena using viscoelastic models.\citep{bazilevskii2005,hoath2012,mcilroy2013,sen2021} In another nozzle-less printing technique called blister-actuated laser induced forward transfer (BA-LIFT), Turkoz et al.\citep{turkoz2018} experimentally investigated the conditions under which a single droplet is formed after jet formation and found that the Deborah number, $De$, and the Ohnesorge number, $Oh$, were used to predict droplet formation. Here, $De = \lambda/ t_c$, where $\lambda$ is the liquid relaxation time and $t_c$ is the characteristic time scale.

Although there is a growing number of research on inkjet printing of viscoelastic liquids,\cite{BOUSFIELD1986,hoath2009,turkoz2021,torres2022} little is known about the influence of viscoelasticity in the inkjet technology of impulsively-induced focused jets. So far, Onuki et al.\citep{onuki2018} has investigated the low viscoelasticity regime and found that the theory of jet velocity in Newtonian fluids is applicable to low viscoelastic liquids. Franco-Gomez et al.\citep{franco2021} investigated highly viscoelastic jet behavior and found an increase in the maximum jet velocity due to elasticity. However, little physical understanding of the effect of viscoelasticity on the behavior of focused jets has been achieved in either of these investigations. Since a impulsively-induced focused jet technology can generate jets at high velocities,\citep{onuki2018} it is possible to investigate viscoelastic jet behavior at higher velocities than with conventional inkjet technology.

In this paper, we quantify the effect of viscoelasticity on pinch-off behavior at high velocities using impulsively induced focused jets. To this purpose, a systematic investigation is conducted by varying the magnitude of viscoelasticity over a wide range. The formation of viscoelastic jets is observed in detail with a high-speed camera, and the jet behavior is classified into phase diagrams using the Reynolds number, $Re$, and the Weissenberg number, $Wi$. Here $Re = \rho U_0 r/ \mu$, where $U_0$ is the characteristic velocity, and $Wi = \lambda U_0 / r$. The reason for adapting $Re$ and $Wi$ is discussed in ``Results and Discussion" section. Furthermore, we will investigate whether the models proposed in DoD printing \citep{hoath2012,mcilroy2013} can be applied to viscoelastic focusing jets, and attempt to provide a physical interpretation of viscoelastic jet behavior.

\section{Experimental Section}
\subsection{Impulsively-induced focused jet generation}
We use two different methods to vary the initial velocity of the fluid while considering a wide range of parameters. The first method generates jet using a test tube. This method allow us to generate highly controllable focused jet easily and covers the range of parameters within $0.06 < Wi < 120$ and $16 < Re < 5900$. The second method uses a double-layer narrow tube to generate jet that is capable of ejecting liquid with high elastisity, and also consider a wide parametric range that covers $3.3 < Wi < 16000$ and $24 < Re < 2700$. 

\subsubsection{Method I: Test tube impacting on a floor}\label{sec2}
\begin{figure}[t]
    \centering
        \includegraphics[width=\columnwidth]{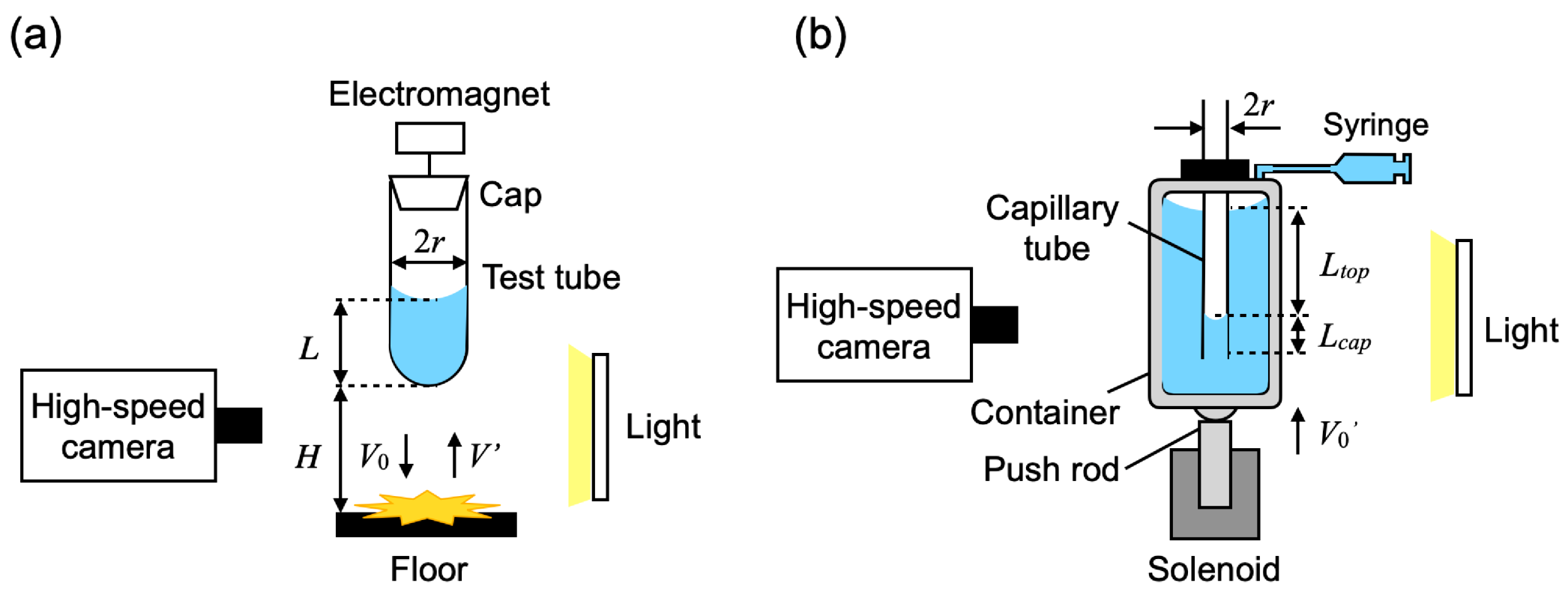}
    \caption{The experimental setup for generating a focused jet using (a) a falling test tube, and (b) double-layer narrow tube, where the capillary tube placed inside the container with the liquid level inside the tube set deeper than that of outside the tube.}
    \label{fig:fig1}
\end{figure}

A schematic of the experimental setup for method I is shown in Figure \ref{fig:fig1}a. A test tube (A-16.5, Maruemu, inner diameter $2r = 14.3$ mm) is filled with liquid (liquid height, $L = 32$ mm) and suspended by an electromagnet at a dropping height, $H$. When the electromagnet is switched off, the test tube falls freely. When the test tube collides with the floor, the liquid is accelerated over a very short time, and as a result, an upward oriented focused liquid jet is formed from the gas-liquid interface.\citep{kamamoto2021, Yukisada2018} 

In this experiment, the initial velocity of the fluid $U_0$ is controlled by varying the drop height of the test tube ($H$ = 3--300 mm) and the floor material (SS400, nitrile rubber). $U_0$ is the velocity of the gas-liquid interface just after the impact, and is the sum of the falling velocity of the test tube (velocity just before impact) $V_0$ and the rebound velocity relative to the floor $V^{\prime}$, expressed as in $U_0 = V_0 + V^{\prime}$.\citep{kiyama2016} In this case, $V_0$ is given by $V_0 = \sqrt{2gH}$, which is derived from the energy conservation law, where $g$ is the gravitational acceleration. $V^{\prime}$ is given by $V^{\prime} = H^{\prime} / t$, where $H^{\prime}$ is the rebound velocity of the test tube $10$ ms after impact ($t$ = 0 ms). 
The jet formation is captured using a high-speed camera (FASTCAM SA-X, Photron) and a backlight (White Led Backlight, Phlox) at 30,000 fps.

\subsubsection{Method II: Double-layer narrow tube}

A schematic of the experimental setup for method II is shown in Figure \ref{fig:fig1}b. We use an ejection device with a capillary tube to generate jets using the impulsive force.\citep{onuki2018} 
In this experiment, a glass capillary tube (Fuji Science Industry, inner diameter $2r = 2$ mm) is inserted into a container partially filled with liquid and sealed outside the tube. We control the pressure of the sealed air outside the tube using a syringe in order to push the liquid interface inside the capillary near to the bottom of the container and thus change the position of the liquid surface inside the capillary $L_{cap}$. The liquid level inside the tube is kept deeper than that outside the tube.
An electromechanical device (ShinDengen solenoids, KGK power supply) is used to apply a vertical upward impulsive force at a velocity $V_0^{\prime}$. The liquid inside the tube is rapidly accelerated and a focused jet is generated from the inside of the tube. The initial velocity $U_0$ of the generated jet varies depending on the ratio of the distance between the gas-liquid interface outside the tube $L_{top}$ and the liquid surface position inside the tube $L_{cap}$. The initial velocity $U_0$ is calculated using the following equation:\citep{onuki2018}

\begin{equation}
U_0 = \frac{L_{top}}{L_{cap}}V_0^{\prime}.
\label{eq4}
\end{equation}

\noindent
The jet formation is captured using a high-speed camera and a backlight at 30,000--60,000 fps. 


\subsection{Properties of the liquid solutions}
Polymer solutions were used as the viscoelastic liquids in the experiments (Table \ref{table}). We used two types of polymers: polyethylene oxide (Sigma-Aldrich, molecular weight $M_w$ = 0.4, 2, 5, and 8 M (= million), polymer concentration $c$ = 0.5, 1, and 2 wt$\%$) and polyacrylamide (Sigma-Aldrich, $M_w$ = 5--6 M, $c$ = 0.2, 0.4, and 0.8 wt$\%$). The solutions were prepared by adding polyethylene oxide (PEO) and polyacrylamide (PAM) to pure water and aqueous-glycerin solution (glycerol: 70 wt\%, water: 30 wt\%, G70W30) as solvents and stirring at 650 rpm, 60 $^{\circ}$C for 24 to 48 hours.


\begin{table}[t]
\caption{The physical properties of the solutions}\label{table}
\begin{tabular*}{\textwidth}{@{\extracolsep\fill}llrrr}
\hline
 & \multirow{2}{*}{Solutions} & Density $\rho$ & Surface tension & Relaxation time \\\
 &  & [kg/m$^3$] & $\sigma$ [mN/m] & $\lambda$ [ms]  \\
\hline
\multirow{6}{*}{PEO} & 0.4M 1wt\% & 999.7 & 64.0 & 1.9 \\
& 2M 1wt\% & 998.3 & 64.1 & 12.4 \\
& 5M 0.5wt\% & 996.4 & 62.8 & 35.4 \\
& 5M 1wt\% & 998.0 & 62.0 & 57.2 \\
& 8M 1wt\% & 981.4 & 60.6 & 207.6 \\
& 8M 2wt\% & 999.1 & 62.1 & 301.4 \\
\hline
\multirow{4}{*}{PAM} & 5-6M 0.2wt\% & 995.2 & 72.5 & 3.3 \\
& 5-6M 0.4wt\% & 1002.8 & 63.0 & 4.6 \\
& 5-6M 0.8wt\% & 1000.0 & 71.0 & 16.5 \\
& 5-6M 0.4wt\% (G70W30) & 1176.9 & 66.4 & 91.0 \\
\hline
\end{tabular*}
\end{table}

\begin{figure}[t]
    \centering
        \includegraphics[width=0.7\columnwidth]{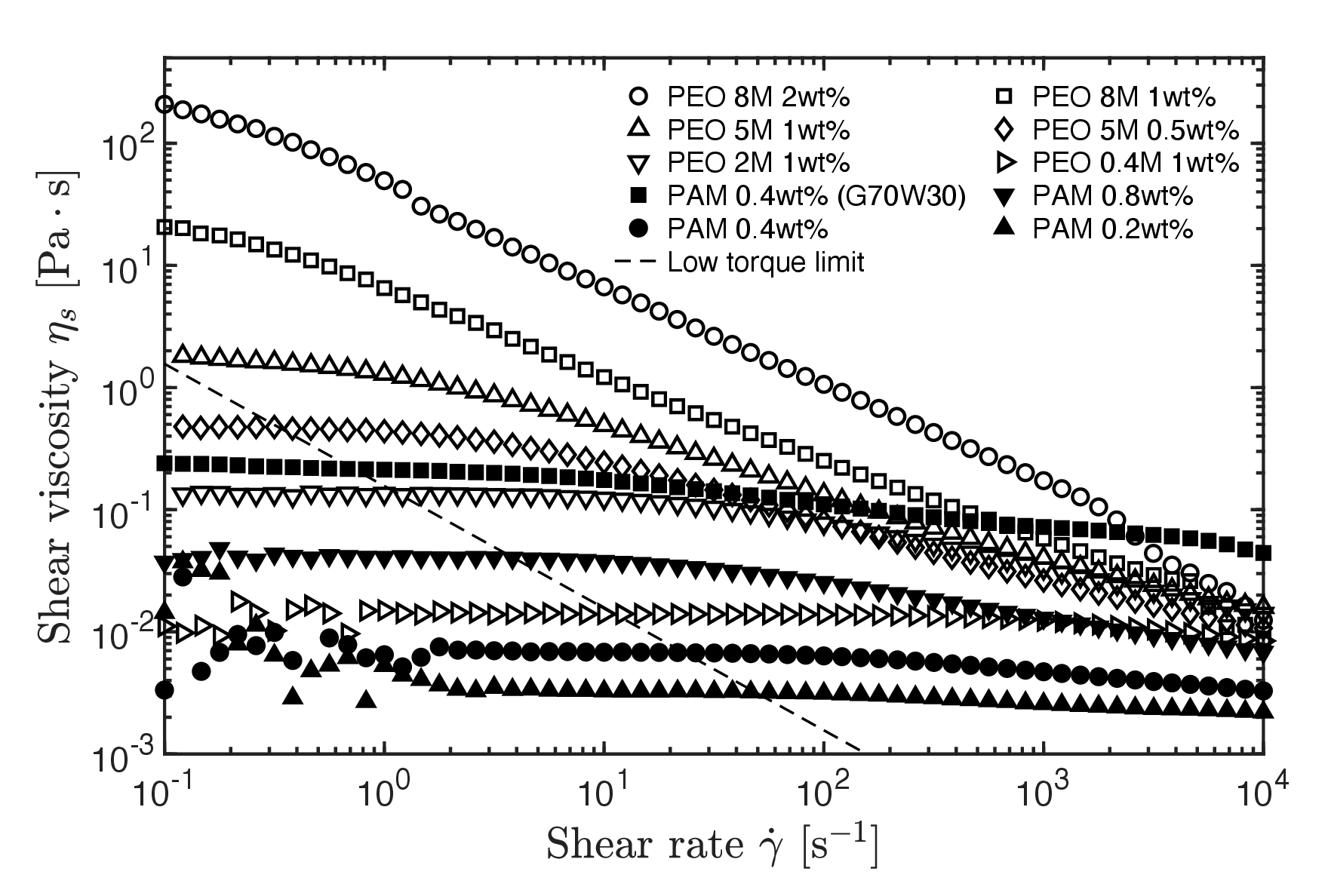}
    \caption{The shear viscosity for the PEO and PAM solutions.}
    \label{fig:fig2}
\end{figure}

Polymer solutions generally exhibit shear-thinning behavior, in which the viscosity decreases with an increasing strain rate. The shear viscosities of the PEO and PAM solutions as measured with a shear rheometer (MCR-302, Anton Paar) are shown in Figure \ref{fig:fig2}. The dashed line in Figure \ref{fig:fig2} represents the low torque limit of the measured shear viscosity due to the torque sensitivity of the shear rheometer.\citep{gaillard2019}
From Figure \ref{fig:fig2}, the shear viscosity $\mu_s$ of both the PEO and PAM solutions decreases with an increasing shear rate $\dot \gamma$, indicating shear-thinning properties. In both solutions, the higher the molecular weight and concentration, the greater the shear viscosity and the greater the slope of the shear curve. The higher the molecular weight and concentration, the greater the viscoelasticity and thus, the stronger the shear-thinning property.\citep{dinic2017}

\begin{figure}[t]
    \centering
        \includegraphics[width=\columnwidth]{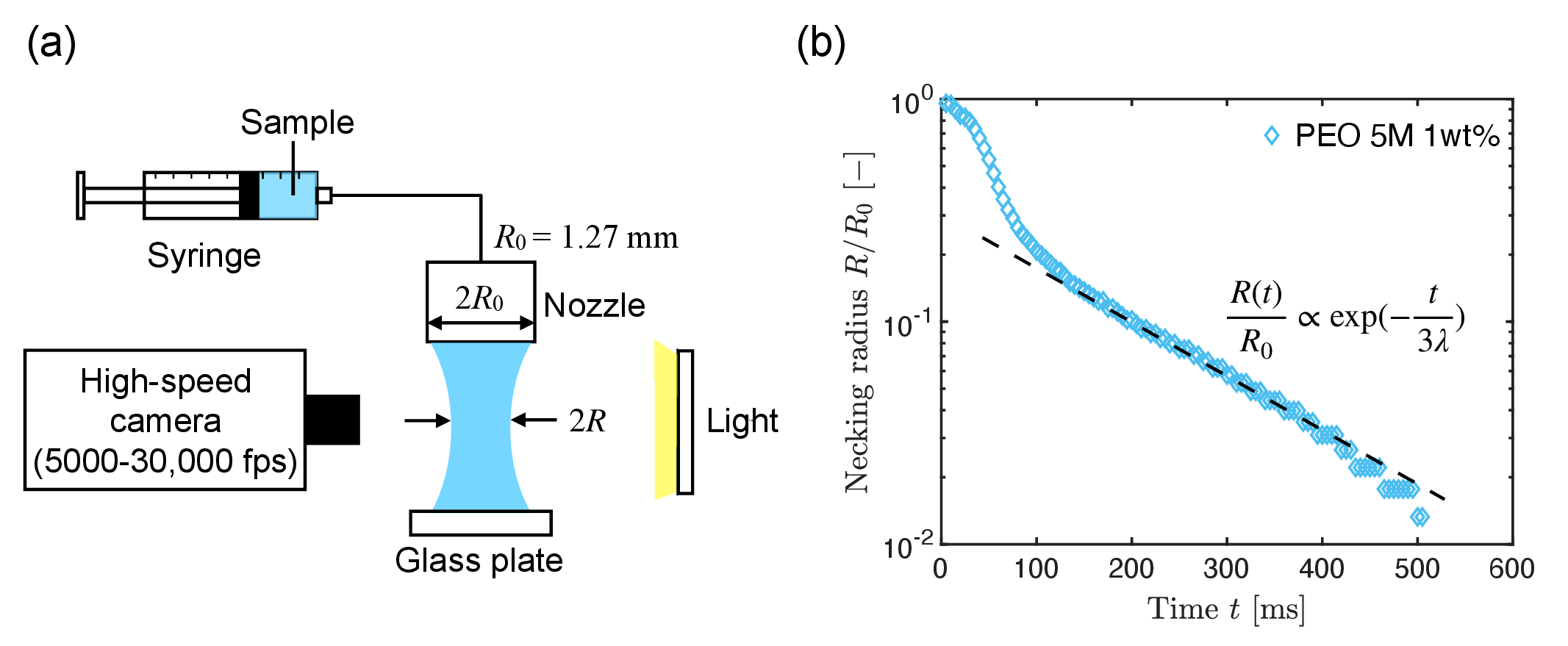}
    \caption{(a) The experimental setup for the dripping-onto-substrate capillary breakup extensional rheometry: DoS-CaBER. (b) The non-dimensional radius (the ratio of the neck radius to the nozzle radius) as a function of time for 5M 1wt\% PEO is shown on a semi-log plot. The dashed line\st{s} shows the fitted region obtained using Eq. \ref{eq5}.}
    \label{fig:fig3}
\end{figure}

Viscoelastic liquids also have spring-like elastic properties. We use the relaxation time $\lambda$ as a physical property to express the magnitude of the elasticity. 
We estimate $\lambda$ using dripping-onto-substrate capillary breakup elongation rheometry (DoS-CaBER), which is a method for evaluating the extensional rheology of viscoelastic liquids.\citep{dinic2017_2} This method enables the measurement of the relaxation times of low-viscoelastic liquids, which are difficult to measure with rheometers.\citep{dinic2015, sur2018} Figure \ref{fig:fig3}a shows the experimental DoS-CaBER setup. First, the sample is pushed out of a syringe to form a droplet at the tip of the nozzle (outer diameter $2 R_0 = 1.27$ mm, inner diameter $2 R_i = 0.97$ mm). When the droplet adheres to the glass plate with its own weight, the liquid spreads on the glass plate and forms a liquid thread due to the surface tension. The liquid thread becomes thin and shreds gradually. The force that resists the surface tension acting on the liquid thread differs depending on the solution, and the elastic force is dominant in viscoelastic liquids (in the elasto-capillary regime).\citep{anna2001} The time evolution equation of the liquid thread radius $R$ formed in this case is expressed as

\begin{equation}
\frac{R(t)}{R_0}{\propto}\exp(-\frac{t}{3\lambda}).
\label{eq5}
\end{equation}

\noindent
where $R_0$ is the nozzle radius and $t$ is the time since the liquid thread was formed.\citep{mathues2018} Therefore, the relaxation time $\lambda$ of the sample can be calculated from the time evolution of the liquid thread radius $R$.

An example analysis of the liquid relaxation time $\lambda$ using DoS-CaBER is shown in Figure \ref{fig:fig3}b. The horizontal axis is the time $t$ and the vertical axis is the dimensionless liquid thread radius $R/R_0$, which is the liquid thread radius $R$ divided by the nozzle radius $R_0$. The plot shows a 5M 1wt$\%$ PEO solution. Since the extension rate is constant in the elasto-capillary regime,\citep{Entov1997} the extension rate is calculated from the DoS-CaBER experimental results and the elasto-capillary regime is determined. An exponential approximation is made in this region, and the relaxation time $\lambda$ is derived using Eq \ref{eq5}.
relaxation time $\lambda$ is derived using Eq. \ref{eq5}.

The relaxation times $\lambda$ for the PEO and PAM solutions are calculated using the same procedure and the results are shown in Table \ref{table}. The relaxation time $\lambda$ increases with an increasing concentration $c$ and molecular weight $M_w$ for both the PEO and PAM solutions. 
Our results are consistent with previous studies.\citep{dinic2017}
In addition, when comparing 2M 1wt\% and 5M 0.5wt\% PEO solutions, $\lambda$ is larger for 5M 0.5wt\%. This is because the molecular weight $M_w$ makes a larger contribution to the relaxation time $\lambda$ than the concentration $c$ .\citep{bazilevskii2005}

\section{Results and Discussion}\label{sec3}
In our experiment, we have identified two different types of jets through systematic experiments by varying the magnitude of the viscoelasticity and initial velocity.  

\begin{figure}[t]
    \centering
        \includegraphics[width=0.7\columnwidth]{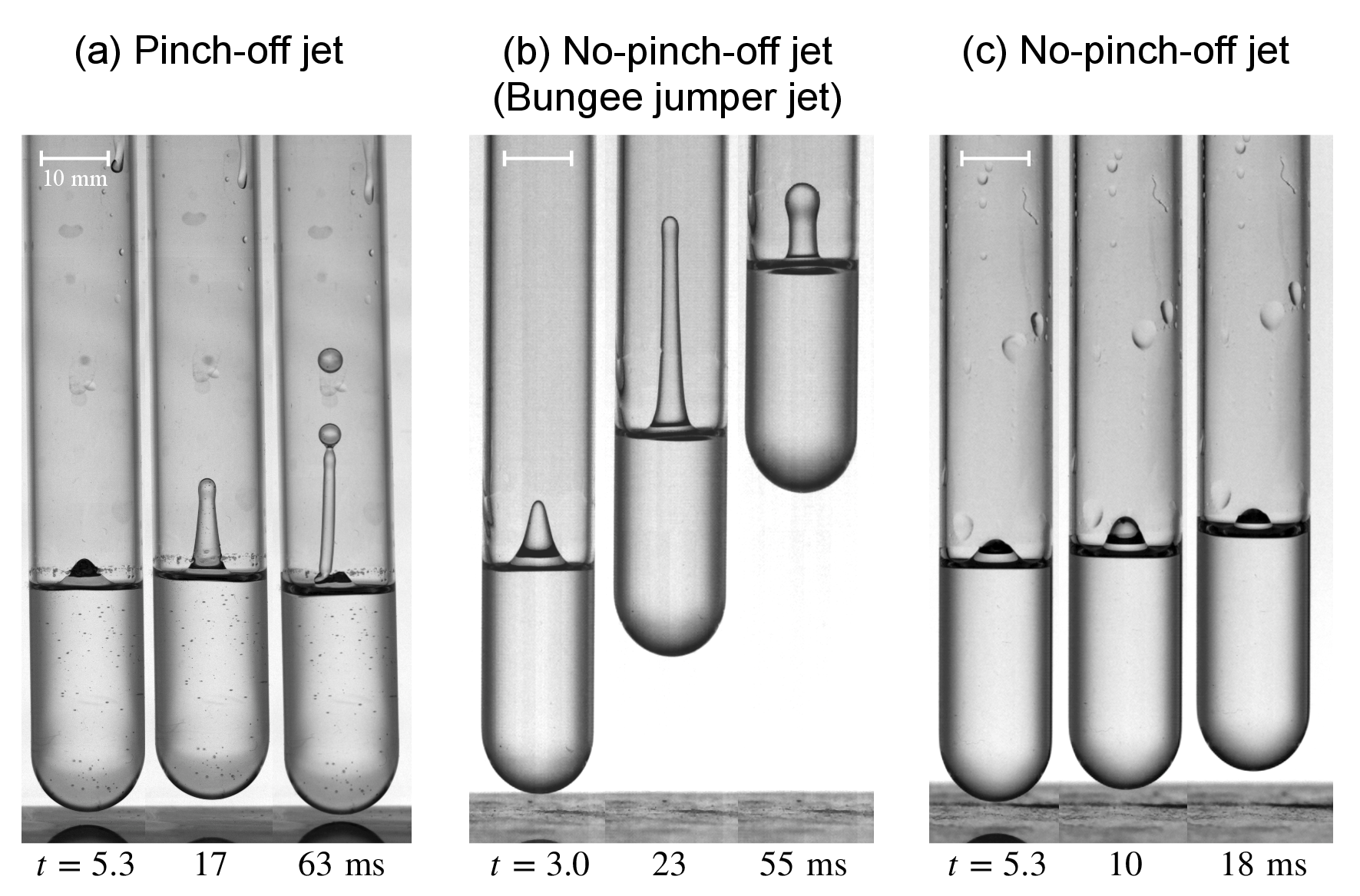}
    \caption{Viscoelastic liquid jets can be divided into two typical types: (a) pinch-off jets and (b)(c) no-pinch-off jets. (a) Images of a jet generated by the ejection of a 0.4M 1wt\% PEO solution at an initial velocity of $U_0 = 0.84$ m/s corresponding to the Supporting Information 1. The jet is ejected from the interface and extends vertically upward, and the tip is pinched off after a certain time. (b) Images of a jet generated by the ejection of an 5M 1wt\% PEO solution at an initial velocity of $U_0 = 2.9$ m/s corresponding to the Supporting Information 2. The jet is elongated after ejection from the interface and, after maximum elongation, pulled back to the initial interface. (c) Images of a jet generated by the ejection of a 8M 1wt\% PEO solution at an initial velocity of $U_0 = 1.2$ m/s corresponding to the Supporting Information 3. The jet is ejected from the interface but returns to the initial interface without elongation.}
    \label{fig:fig4}
\end{figure}

The typical behaviors of the generated viscoelastic liquid jets are shown in Figure \ref{fig:fig4}. Note that the time of the test tube impacting the floor is used as the reference time ($t = 0$ ms), and the time evolution of the jet from that point is shown in Figure \ref{fig:fig4}.
Figure \ref{fig:fig4}a shows images of a jet generated by the ejection of a 0.4M 1wt\% PEO solution at an initial velocity of $U_0 = 0.84$ m/s. The jet ejected from the interface has a shape in which the tip diameter is smaller than the tail diameter, indicating that it is a focused jet. After ejection, the focused jet is elongated vertically upward, and the tip is pinched off at $t=63$ ms. Thus, we define this type of jet as a `pinch-off jet'. Figure \ref{fig:fig4}b shows images of a jet generated by ejecting a 5M 1 wt\% PEO solution at an initial velocity of $U_0 = 2.9$ m/s. The focused jet is elongated after ejection from the interface and reaches its maximum length $32$ mm at $t=23$ ms. After that, the jet returns to the initial interface. It should be noted that the jet is prevented from being pinched off despite the large impact force applied in Figure \ref{fig:fig4}b compared to that in Figure \ref{fig:fig4}a. This phenomenon was also observed in a previous study as a `bungee-jumper jet' when the elasticity of the solution is large,\citep{franco2021} and it is believed that the elastic force contributes to the pulling back of the jet. We define such a jet that pulls back to the initial interface after maximum elongation as a `no-pinch-off jet (bungee-jumper jet)'.
Finally, Figure \ref{fig:fig4}c shows images of a jet generated by ejecting an 8M 1 wt\% PEO solution at an initial velocity of $U_0 = 1.2$ m/s. The jet is ejected from the interface but returns to the initial interface without elongation. The reason for this is presumed to be that the initial velocity is smaller in Figure \ref{fig:fig4}c than in Figure \ref{fig:fig4}b, i.e., the jet is not given a sufficient impulsive force to generate a jet. Such jets are also defined as `no-pinch-off jets'.

These changes in jet behavior are considered to be caused by the balance between the inertial force and the viscoelasticity of the liquid that dominates during jet extension. We organize our results in terms of two dimensionless numbers. 
One is the Reynolds number $Re$, which represents the ratio of the inertial force to the viscous force, and is defined as

\begin{equation}
Re =\frac{\rho U_0 r}{\mu_s},
\label{eq7}
\end{equation}

\noindent
where $\rho$ is the density of the solution. Since the polymer solution used in this experiment has shear-thinning properties, we calculate the strain rate $\dot{\gamma}$ from the elongation behavior of the jet ($\dot{\gamma} = O(10^2$-$10^3)$) and assume the shear viscosity $\mu_s$ at that strain rate as the representative viscosity.
The other is the Weisenberg number $Wi$, which represents the ratio of the elastic force to the viscous force, and is defined as

\begin{equation}
Wi =\frac{\lambda U_0}{r}.
\label{eq6}
\end{equation}

\begin{figure}[h]
    \centering
        \includegraphics[width=0.8\columnwidth]{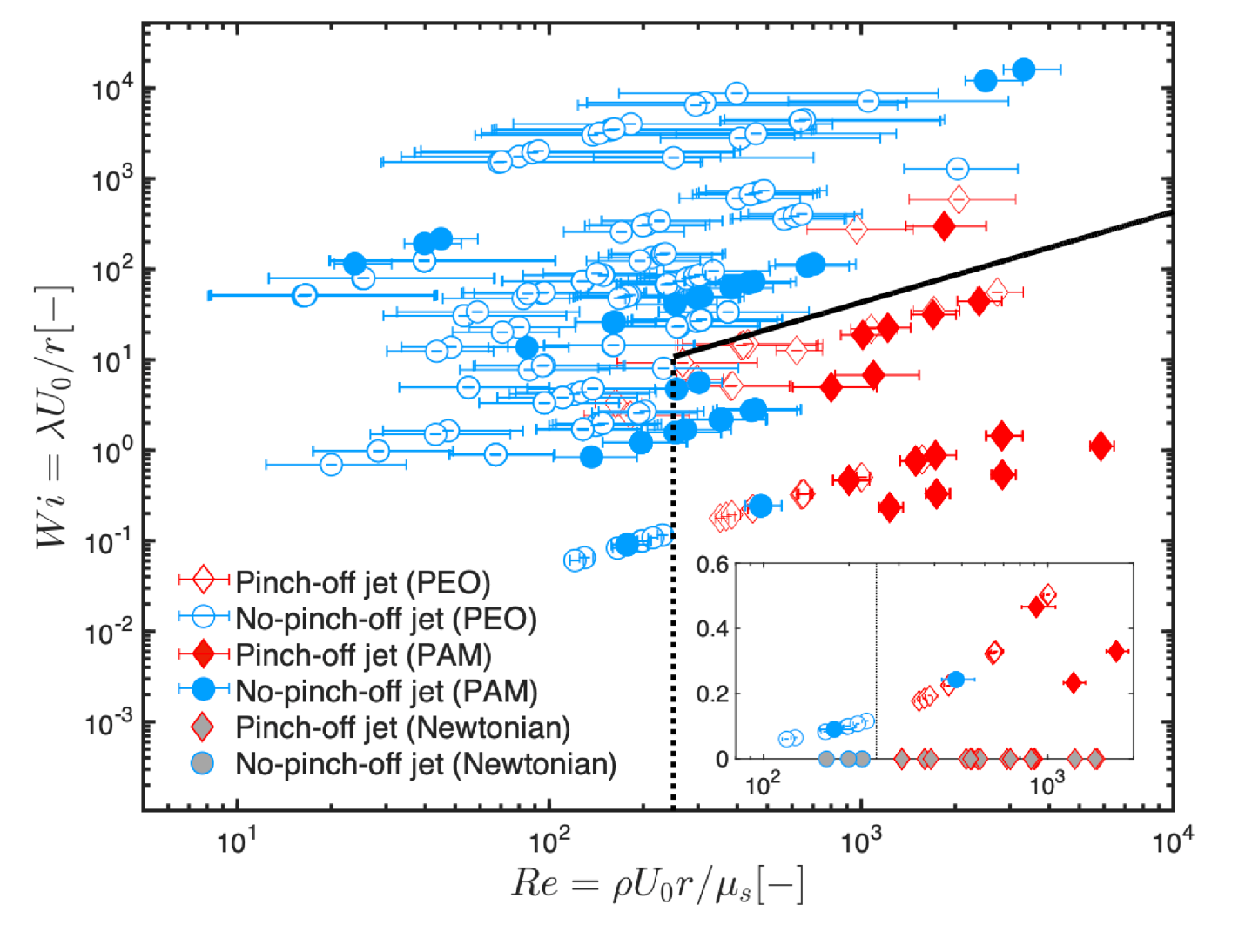}
    \caption{A phase diagram of the impulsively-induced jet behavior of viscoelastic liquids with a Weissenberg number $Wi$ and Reynolds number $Re$. The outer frame of the plot indicates the jet type: red indicates a pinch-off jet and blue indicates a no-pinch-off jet. The markers in the plot indicate the type of solution: an open markers indicates PEO solutions, markers filled with blue and red indicate PAM solutions, and gray markers indicate an aqueous-glycerin solution, which is a Newtonian fluid (See Figure \ref{fig:fig7} for the jet behavior of the Newtonian liquid). The inset shows the region with low $Wi$ ($Wi \leq 0.6$).}
    \label{fig:fig5}
\end{figure}

\noindent
We adapt $Re$ and $Wi$ instead of other dimensionless numbers because they are directly derived from the equations of motion of viscoeastic jet discussed later in this paper.

The results of organizing the jet behaviors using the Reynolds number $Re$ and the Weisenberg number $Wi$ are shown in Figure \ref{fig:fig5}. The error bars in Figure \ref{fig:fig5} indicate the error range associated with the estimation of the shear viscosity $\mu_s$. The higher the viscoelasticity of the liquid, the greater the shear-thinning property, resulting in a larger error especially in the region of high $Wi$.

Figure \ref{fig:fig5} shows that the jet behavior can be divided into two regions, pinch-off, and no-pinch-off jets, depending on $Re$ and $Wi$. The fact that the results can be organized by parameters related to the initial conditions of jet ejection and the physical properties of the solution suggests that these parameters have a significant effect on the behavior of the viscoelastic liquid jets. Furthermore, the experimental results are distributed in a unified manner regardless of the type of polymer (PEO or PAM). Therefore, we believe the effect of the polymer type on the jet behavior is small.
In the high $Wi$ region, the elasticity of the solution affects the jet behavior, but in the low $Wi$ region, it has little effect on the jet behavior and the jet is considered to behave like a Newtonian fluid. Thus, in the following, we will discuss the jet behavior in detail in two regions: 1) the region with $Wi \gtrsim 10$ and 2) the region with $Wi \lesssim 10$.

\subsection{The region with $Wi \gtrsim 10$ (Elastic dominated)}

Figure \ref{fig:fig5} shows that in the region with $Wi \gtrsim 10$, a no-pinch-off jet occurs in the region with high $Wi$, and a pinch-off jet occurs in the region with low $Wi$. It should be noted that even for a sufficiently high $Re$, $Re \gtrsim O(10^3)$, the pinch-off of the jet is inhibited in the region with high $Wi$ ($Wi \gtrsim O(10^3)$). The pinch-off phenomenon is suppressed even in this region, which is a region of high velocity and high viscoelasticity ompared to the range of previous studies on inkjet of viscoelastic liquids.\citep{morrison2010, hoath2009} The elasticity increases with increasing $Wi$, suggesting that bungee-jumper jets are caused by the elasticity of the viscoelastic liquids. Thus, remarkably, we find that the pinch-off is suppressed even in the high-velocity regimes when compared to the range of previous studies on inkjet applications using viscoelastic liquids\citep{morrison2010, hoath2009}.

To understand this phenomenon, we model a jet using a viscoelastic model following previous studies.\citep{mcilroy2013,hoath2012} The effect of elasticity on the jet behavior can be evaluated by introducing a constitutive equation that accounts for elastic stress.\citep{turkoz2021} There are various models, but here we use the finitely extensible non-linear elastic dumbbell model with the Chilcott–Rallison closure approximation, known as the FENE-CR model.\citep{chilcott1988} The FENE-CR model is one of the simple constitutive equations used for the rheological properties of dilute polymer solutions.

Assuming an incompressible fluid, the continuity equation and Navier--Stokes equation can be expressed using the fluid velocity $\mathbf{u}$, pressure $p$, and stress tensor $\boldsymbol{\sigma}$ as

\begin{equation}
\nabla \cdot \mathbf{u}  =0,
\label{eq8}
\end{equation}

\begin{equation}
\rho \frac{D \mathbf{u}}{Dt} = - \nabla p + \nabla \cdot \boldsymbol{\sigma}.
\label{eq9}
\end{equation}

\noindent
In the FENE-CR model, the stress is given by

\begin{equation}
\boldsymbol{\sigma} = 2 \mu_s \mathbf{E} +Gf(\mathbf{A-I}),
\label{eq10}
\end{equation}

\noindent
where $\mathbf{E} = (\nabla \mathbf{u} + \nabla \mathbf{u} \rm{^T}$)/2 is the strain rate tensor, $G$ is the elastic modulus, $\mathbf{I}$ is the unit matrix, $\mathbf{A}$ is the conformation tensor related to the deformation of the polymer chains, and $f$ is a nonlinear function that enforces finite extensibility. We note

\begin{equation}
\frac{D \mathbf{A}}{Dt} = \nabla \mathbf{u} \rm{^T} \cdot \mathbf{A} + \mathbf{A} \cdot \nabla \mathbf{u} -\mathit{\frac{f}{\lambda}}(\mathbf{A-I}).
\label{eq11}
\end{equation}

\noindent
Using the elongation limit of the polymer chain $L$, $f$ is expressed as

\begin{equation}
f = (1-tr(\mathbf{A})/L^2)^{-1} .
\label{eq12}
\end{equation}

\begin{figure}[b]
    \centering
        \includegraphics[width=0.7\columnwidth]{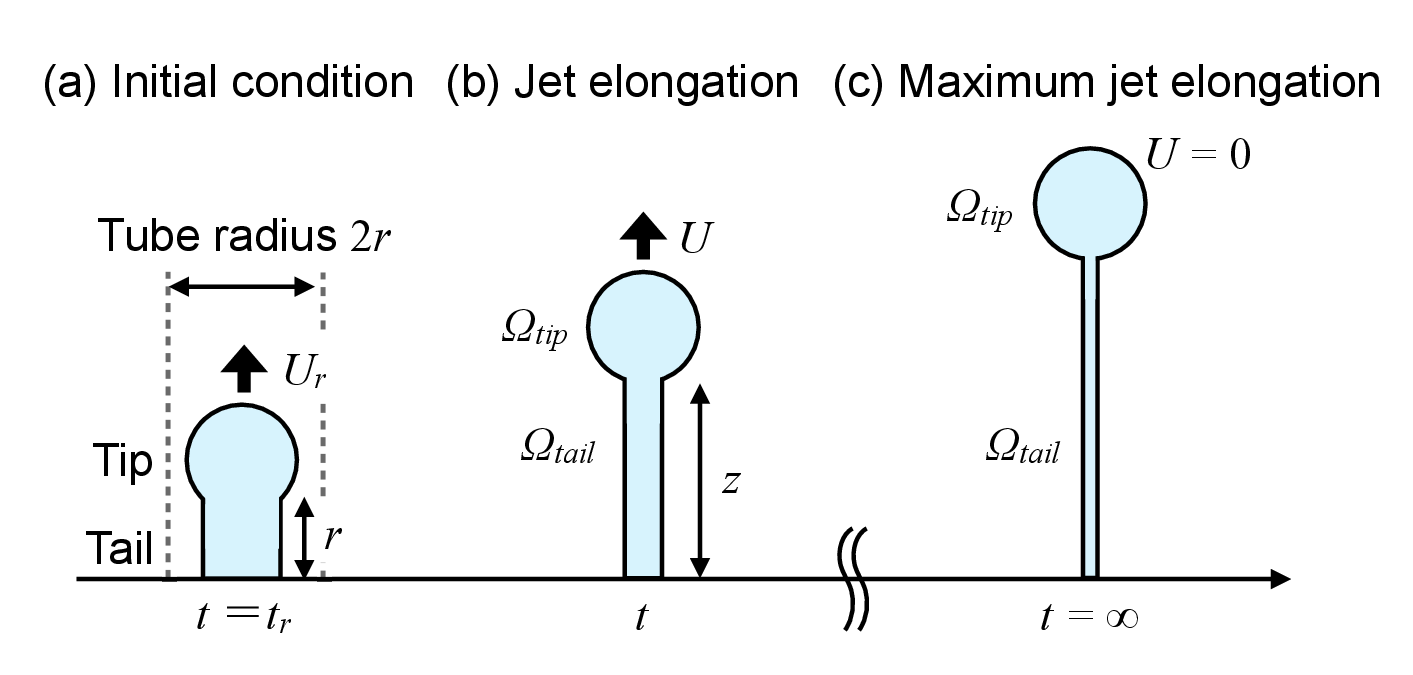}
    \caption{A simplified model of the viscoelastic jet.}
    \label{fig:fig6}
\end{figure}

Since the tip of the jet is faster than the tail, we assume that the inertia at the tip of the jet affects the jet behavior, and propose a simple model as shown in Figure \ref{fig:fig6}. This model is based on the model outlined by previous studies.\citep{hoath2012, mcilroy2013}. Hoath et al.\citep{hoath2012} predicted from modeling that the maximum polymer concentration at which a jet of a certain speed can be formed scales with molecular weight, and demonstrated agreement with experimental data. Later, McIlroy et al.\citep{mcilroy2013} validated the predictions of the simple model of jets given by Hoath et al.\citep{hoath2012} against both experimental observations and numerical simulations and were able to identify all three asymptotic regions identified by Hoath et al.\citep{hoath2012} for low viscosity solvents. We define the initial condition to be $t = t_r$, where $t_r$ is the time when the jet tip is ejected by a distance $r$ (=tube radius) from the initial interface in Figure \ref{fig:fig6}a. We set the initial condition to a value used by previous study.\cite{mcilroy2013}
The jet velocity at this time is $U_r$. The jet then decelerates to a velocity $U$, and the jet tail deforms uniformly with time from an initial length $r$ to $z$ at time $t$ (Figure \ref{fig:fig6}b). The volume of fluid at the tip $\Omega_{tip}$ and at the tail $\Omega_{tail}$ are assumed to remain constant. 

For this simple model, the velocity $U$ is given by $U = dz/dt$. Assuming that the only forces acting on the jet tip are from the stress difference in the tail\st{,}. The equation of motion of the tip is given by

\begin{equation}
\rho \Omega_{tip} \frac{dU}{dt} = -\frac{\Omega_{tail}}{z}\bigg(\frac{\mu_e U}{z} + Gf(A_{zz}-A_{rr})\bigg),
\label{eq13}
\end{equation}

\noindent
where $\mu_e$ is extensional viscosity.\citep{mcilroy2013} Assuming an axisymmetric flow, Eq. \ref{eq11} can be expanded into an axial component $A_{zz}$ and a radial component $A_{rr}$, and the orthogonal stresses $A_{rz}$ and $A_{zr}$ in the jet are neglected ($A_{rz} = A_{zr} = 0$). Thus, $A_{zz}$ and $A_{rr}$ can be expressed as follows:

\begin{equation}
\frac{dA_{zz}}{dt} = \bigg(\frac{2U}{z}-\frac{f}{\lambda}\bigg)A_{zz} + \frac{f}{\lambda},
\label{eq14}
\end{equation}

\begin{equation}
\frac{dA_{rr}}{dt} = -\bigg(\frac{U}{z}+\frac{f}{\lambda}\bigg)A_{rr} + \frac{f}{\lambda}.
\label{eq15}
\end{equation}

\noindent
Furthermore, assuming that the axial elongation of the polymer chain is sufficiently large ($A_{zz}\gg 1$ → $f \sim 1$, $A_{zz} \gg A_{rr}$), we can write $ A_{zz} \sim (z/r)^2e^{-t/\lambda}$ and $A_{rr} \sim 0$. 
Therefore, the deceleration $\Delta U$ at position $z$ ($z > r$) is obtained by integrating Eq. \ref{eq13} at time $t$ (Figure \ref{fig:fig6}b).

\begin{equation}
\rho \Omega_{tip} \mathit{\Delta} U(z) = -\Omega_{tail} \bigg(\mu_e \bigg(\frac{1}{r} - \frac{1}{z}\bigg) + \frac{G}{r^2}\int_{0}^{t}ze^{-\frac{t}{\lambda}}dt\bigg).
\label{eq16}
\end{equation}

\noindent
By assuming that $\Delta U/U$ is small, $z = r + U_rt$, and taking the limit $t \rightarrow \infty$ following previous study,\citep{hoath2012} we can approximate the integral as

\begin{equation}
\int_{0}^{\infty}ze^{-\frac{t}{\lambda}}dt = \int_{0}^{\infty}(r + U_rt)e^{-\frac{t}{\lambda}}dt \sim U_r \lambda^2.
\label{eq_integral}
\end{equation}

\noindent
In addition, assuming $U = 0$ when $t \rightarrow \infty$ (Figure \ref{fig:fig6}c), we get

\begin{equation}
\rho \Omega_{tip}U_r = \Omega_{tail}{ }\bigg(\frac{\mu_e}{r} + \frac{U_r}{r^2}G\lambda^2\bigg).
\label{eq17}
\end{equation}

\noindent
When the elastic term is sufficiently large compared with the viscous term, Eq. \ref{eq17} can be transformed into

\begin{equation}
\rho \Omega_{tip}U_r = \Omega_{tail} \frac{U_r}{r^2}G\lambda^2.
\label{eq18}
\end{equation}

\noindent
Assuming that the jet tip diameter $d_{tip}$ and the tail diameter $d_{tail}$ satisfy $d_{tip} \sim \alpha r$ and $d_{tail} \sim r$ in the initial state (Figure \ref{fig:fig6}a), then
$\Omega_{tip} \sim 4\pi(\alpha r/2)^3/3$, $\Omega_{tail} \sim \pi r^3/4$. We note that $\alpha$ is a constant.
Furthermore, assuming $G \sim \mu_s / \lambda$ and $U_r \propto U_0$, and substituting these into Eq. \ref{eq18}, we obtain

\begin{equation}
Re \sim \frac{3}{2 \alpha^3} Wi.
\label{eq19}
\end{equation}

\noindent
For $Re \lesssim 3/(2 \alpha^3) Wi$, the jet is pulled back by the elastic force, resulting in a no-pinch-off jet, and for $Re \gtrsim 3/(2 \alpha^3) Wi$, a pinch-off jet occurs due to the lack of elastic force.
The experimental value of $\alpha$ is 0.4.
From the above, we suggest that the threshold for the occurrence of pinch-off jets in the region of $Wi \gtrsim 10$ can be expressed by the following equation:

\begin{equation}
Re \gtrsim 23.4 \, Wi.
\label{eq20}
\end{equation}

\subsection{The region with $Wi \lesssim 10$ (Viscous dominated)}
From Figure \ref{fig:fig5}, no-pinch-off jets are observed for $Re\lesssim 250$ the dashed vertical line, and pinch-off jets are observed for $Re\gtrsim 250$ when $Wi \lesssim 10$. This is considered to be because the elasticity of the solution has little effect on the jet behavior in the low $Wi$ region, and the jet behaves like a Newtonian fluid. In a previous study, the effect of viscosity on the jet velocity of a Newtonian fluid was investigated.\citep{onuki2018} It was reported that a viscous boundary layer develops on the wall surface at $Re \lesssim 200$, which prevents the focusing effect of the flow. Therefore, the lower $Re$ is in the present study, the more the viscous boundary layer develops on the wall surface and near the free surface, which prevents jet extension.

\begin{figure}[t]
    \centering
        \includegraphics[width=0.7\columnwidth]{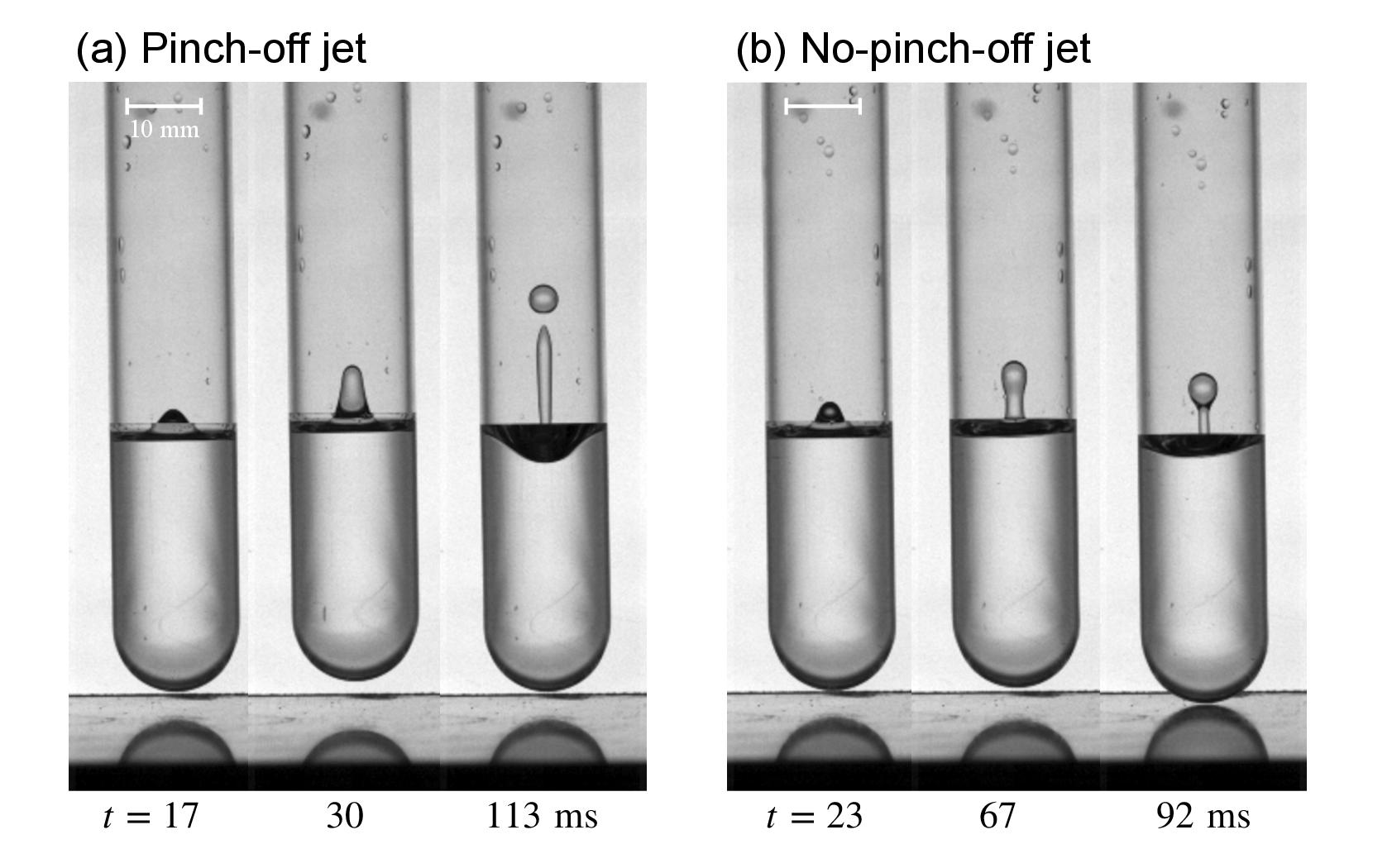}
    \caption{The jet behavior of aqueous-glycerin solution (G70W30). (a) A jet ejected from the interface and extended vertically upward, with the tip pinched off after a certain time when $Re \sim 370$. (b) A jet ejected from the interface but returned to the initial position without being pinched off when $Re \sim 230$.}
    \label{fig:fig7}
\end{figure}

When the fluid behaves in a Newtonian Fluid-like manner ($\mu_e = 3 \mu_s$) and viscosity is dominant, Eq. \ref{eq17} can be transformed into




\begin{equation}
\rho \Omega_{tip}U_r = \Omega_{tail}\frac{3 \mu_s}{r}.
\label{eq21}
\end{equation}

\noindent
Assuming $d_{tip} \propto r$ and $d_{tail} \sim r$ as before, $\Omega_{tip}$ and $\Omega_{tail}$ can be calculated and substituted into Eq. \ref{eq21}. Also, by assuming $U_r \propto U_0$, Eq. \ref{eq21} can be transformed into 

\begin{equation}
Re \sim \beta,
\label{eq22}
\end{equation}

\noindent
where $\beta$ is a constant. For $Re \lesssim \beta$, a no-pinch-off jet is expected to occur because the viscosity prevents the jet extension, and for $Re \gtrsim \beta$, a pinch-off jet is expected to occur because the inertial force is sufficient.

To confirm that the elasticity of the solution does not affect the jet behavior in this viscous dominated region, we investigated the jet behavior of a Newtonian fluid, aqueous-glycerin solution (G70W30) where the density $\rho$, viscosity $\mu_s$, and surface tension $\sigma$ of the glycerin solution are $\rho = 1175.4$ kg/m$^3$, $\mu_s = 0.0169$ Pa$\cdot$s, and $\sigma = 68.4$ mN/m, respectively.
Note that we do not consider the effect of the surface tension on the jet behavior since the surface tension for all viscoelastic and newtonian liquids are similar ($\sigma$ = 62.0-72.5 mN/m).
Figure \ref{fig:fig7} shows a typical image of a G70W30 jet. As shown in Figure \ref{fig:fig7}a, when $Re \sim 370$, the jet is pinched off at the tip after extension and a pinch-off jet is generated. On the other hand, when $Re \sim 230$, as shown in Figure \ref{fig:fig7}b, the jet returns to the initial interface without being pinched off after ejection, so a no-pinch-off jet is generated. We determined from the experiments that the jet behaviors of the Newtonian fluid transitioned from no-pinch-off to pinch-off at $Re \sim 250$ (Figure \ref{fig:fig7}). Therefore, in Eq. \ref{eq22}, the $\beta$ obtained from the experiments is $\beta \sim 250$, and the threshold for the occurrence of pinch-off jets in the region of $Wi \lesssim 10$ is

\begin{equation}
Re \gtrsim 250.
\label{eq23}
\end{equation}

In Figure \ref{fig:fig5}, the no-pinch-off and pinch-off jets can be generally distinguished using Eq. \ref{eq20} (solid line) and Eq. \ref{eq23} (dashed line). Therefore, it is evident that the jet behavior can be classified through modeling with the FENE-CR model.

\section{Conclusions}
In this study, we have conducted systematic experiments on jet behaviors in the high-viscoelasticity and high-velocity regime, which have not been investigated before. We generated a focused jet using polymer solution with a wide range of viscoelasticities and observed two characteristic behaviors: 1) The jet elongates and after a certain amount of time, the tip gets pinched off (pinch-off jet) and 2) The jet returns to the initial interface without pinch-off (no-pinch-off jet). The no-pinch-off jet in the highly elastic region gets pulled back after a large elongation. This phenomenon has been observed as a bungee-jumper jet in previous studies,\citep{morrison2010, hoath2009} but this study shows that the bungee-jumper jet is observed not only in low-velocity regimes but also in the higher velocity and higher elasticity regime. We suggest that these jet behaviors can be reasonably classified using Reynolds number $Re$ and Weisenberg number $Wi$, which are dimensionless numbers composed of the initial conditions of the jet ejection and the rheological properties of the solution. Furthermore, to understand this phenomenon we used a viscoelastic model, the FENE-CR model, and rationalized the experimental results.

We believe that this study provides significant insight into the viscoelasticity of jetswith a focused geometry. This work will be useful for clarifying the effects of viscoelasticity on the breakup processes of various liquid jets in high-velocity regimes.

\begin{acknowledgement}
This work was funded by the Japan Society for the Promotion of Science (Grant Nos. 20H00223, 20H00222, and 20K20972), the Japan Science and Technology Agency PRESTO (Grant No. JPMJPR21O5), and Japan Agency for Medical Research and Development (Grant No. JP22he0422016).
\end{acknowledgement}

\section{Conflict of interest}
The authors declare no conflict of interest.

\bibliography{reference}
\end{document}